\documentstyle[10pt,psfig,aaspp4]{article}

\begin{document}

\title{
Correlation Between Luminosity and Accretion Torque in 4U~1626--67 and
GX~301--2}

\author{
Brian~A.~Vaughan\altaffilmark{1} and  Shunji~Kitamoto\altaffilmark{2,3}}
\altaffiltext{1}{Space Radiation Laboratory, California Institute
of Technology, MC 220--47, Pasadena CA 91125; brian@srl.caltech.edu}
\altaffiltext{2}{Department of Earth and Space Science, 
Graduate School of Science, Osaka University\\
1--1, Machikaneyama-cho, Toyonaka, Osaka, 560, Japan; 
kitamoto@ess.sci.osaka-u.ac.jp}
\altaffiltext{3}{CREST, Japan Science and Technology Corporation (JST)}

\begin{abstract}

We present X-ray light curves and energy spectra  for the persistent
accreting pulsars 4U~1626--67 and GX~301-2 measured by the All-Sky
Monitor (ASM) on {\it Ginga} from 1987 March -- 1991 October.  We compare these
with simultaneous and near simultaneous measurements of spin
frequency and flux by other instruments, principally the Burst and Transient
Source Experiment (BATSE) on the Compton Gamma Ray Observatory (CGRO).  
A dramatic change in the shape of the X-ray spectrum and a $\sim$20\%
decrease in the 1--20 keV X-ray flux accompany the 1990 
transition from steady spin up to steady spin down in 4U~1626--67. The
{\it Ginga} ASM is the only instrument to observe 4U~1626--67 during both spin
up and spin down.  We show that the distance to 4U~1626--67 is
$\gtrsim$5\,kpc.  If 4U~1626--67 is a near-equilibrium rotator and if the
0.04\,Hz Quasi-Period Oscillations seen during spin up are magnetospheric
beat-frequency oscillations, then the distance to the source is $\sim$5\,kpc,
assuming a neutron-star mass of 1.4\,$M_\odot$, radius 10\,km, and
moment of inertia $10^{45}$g\,cm$^2$.

The X-ray flux of GX~301--2 measured with the ASM varies with orbital
phase.  The flux peaks shortly before periastron, with a secondary maximum
near apastron.  Such variations were seen previously in the 20--50\,keV
pulsed flux with BATSE.  The ASM observations confirm that the 20--50\,keV
pulsed flux in GX~301--2 is a good tracer of the bolometric flux.
The X-ray flux in GX~301--2 was a factor of $\sim$2 larger than average
during the periastron passage prior to an episode of persistent
spin up in 1991 July observed with BATSE that lasted 
half an orbit and resembled outbursts seen
in transient X-ray pulsars.  No ASM observations were available during
the spin-up episode.

\end{abstract}

\keywords{
Stars: Binaries: General ---  Stars: Neutron ---- 
Accretion, Accretion Disks --- Stars: Pulsars ---
X-rays: Stars --- Stars: Individual (4U1626-67, GX301-2)}

\section{Introduction}

Accreting pulsars consist of rotating, magnetized neutron stars
that accrete matter from a companion (Pringle \& Rees 1972;
Davidson \& Ostriker 1971).  Material may be captured either from the
stellar wind of the companion (wind accretion) or through Roche-lobe
overflow of the mass donating star (disk accretion).
The flow of material, either radially inward or through an accretion disk,
is interrupted when magnetic stresses dominate material stresses
at the magnetospheric radius, $r_{\rm m}=K\mu^{4/7}(GM_{\rm x})^{-1/7}
\dot M^{-2/7}$.  Here, $\mu$ is the neutron-star magnetic moment, 
$M_{\rm x}$ is its mass, $\dot M$ is the mass accretion rate, and K is
a constant of order unity.  For $K=0.91$, $R_m$ is equal to the Alfven 
radius for spherical accretion.

In the simplest picture of disk
accretion, matter becomes attached to magnetic field lines at $r_{\rm m}$
and transported to the magnetic poles.  If the angular momentum of material
captured from the disk at $r_{\rm m}$ is carried to the neutron star, the
neutron star experiences an accretion torque 

\begin{equation}
N=\dot M\sqrt{GM_{\rm x}r_{\rm m}}.
\label{N}
\end{equation}
A star with moment of inertia $I_{\rm x}$ subject
 to the torque in equation~(\ref{N}) will spin up at a rate

\begin{equation}
2\pi I_{\rm x}\dot\nu = (GM_{\rm x})^{3/7}\mu^{2/7}\dot M^{6/7}.
\label{nudot}
\end{equation}
Assuming the gravitational potential energy of the accreted material is 
converted to X-rays at the neutron-star surface, the X-ray luminosity will
be 

\begin{equation}
L_{\rm x}\approx G\dot M M_{\rm x} / r_{\rm x},
\label{Lx}
\end{equation}
where $r_{\rm x}$ is
the neutron-star radius.  From equations~(\ref{nudot}) and (\ref{Lx}), the rate
of spin up is related to the X-ray intensity through
$\dot\nu\propto L_{\rm x}^{6/7}.$

If the magnetospheric radius lies outside the corotation radius, $r_{\rm co}$,
where the Keplerian orbital frequency equals the spin frequency of the 
neutron star, matter that becomes attached to field lines may be expelled
from the system.  Accretion is then centrifugally inhibited.  In this 
``propellor'' regime, the neutron star may spin down rapidly.  Neutron 
stars accreting at a constant rate thus tend toward an equilibrium spin
period where $r_{\rm m}\approx r_{\rm co}$, given by

\begin{equation}
P_{\rm eq}= 2\pi(GM_{\rm x})^{-5/7}\mu^{6/7}\dot M^{-3/7}.
\label{Peq}
\end{equation}
The relation between torque and luminosity near
equilibrium is expected to be more complicated than equation~(\ref{N}).

Magnetic accretion occurs
in a variety of astrophysical systems, including magnetic CVs and T Tauri
stars (Warner 1990; Konigl et al. 1991).
Accreting pulsars are well suited for studying accretion phenomena.
Their small moments of inertia and strong magnetic fields
result in measureable changes in the spin frequency, $\nu$, in
hours to days with current instruments (e.g. Nagase 1989).
In particular, accreting pulsars open the
possibility of probing the interaction between material in the
accretion disk and the magnetic field through measurements of how
$\dot{\nu}$ depends upon $\dot{M}$.  

A correlation between spin-up rate and X-ray luminosity has been
observed in outbursts of 5 transient systems; between the spin-up rate
and the 1--20 keV flux measured with EXOSAT in 
\hbox{EXO 2030+375} (Parmar, White \& Stella 1989; Parmar et al. 1989;
Reynolds et al. 1996),
and between the spin-up rate and the flux above 20\,keV (in some cases the
pulsed flux)
measured with BATSE in \hbox{2S 1417--62} (Finger, Wilson \& Chakrabarty 1996),
\hbox{A 0535+26} (Bildsten et al. 1997; Finger, Wilson \& Harmon 1996), 
\hbox{GS 0834-43} (Wilson et al. 1997)
and \hbox{GRO J1744--28} (Bildsten et al. 1997).  
All of these outbursts had luminosities and accretion torques
exceding those normally observed in most persistent sources.  They
almost certainly satisfy $r_m\ll r_{co}$ during most of the outburst.

The case for torque-luminosity correlations in
persistent sources is less clear.
The disk-fed systems Cen X-3 and GX 1+4 both
exhibit strong flares lasting several days.  In Cen X-3, no
correlation between the pulse frequency history and the X-ray flux
history has been found (Tsunemi, Kitamoto \& Tamura 1996;
Bildsten et al. 1997). 
In GX 1+4, there is
an {\it anti}correlation between the 20--50 keV pulsed flux
measured with the BATSE and
accretion torque (Chakrabarty et al. 1997b). 
In the wind-fed system GX 301--2,
the X-ray flux has been found to vary with orbital phase.  However,
continuous measurements with BATSE show no correlation between orbital
phase and either the magnitude or the sense of the accretion torque
(Koh et al. 1997).

In this paper we present measurements of X-ray flux for the accreting
pulsars 4U~1626--67 and GX~301--2 made with the
All-Sky Monitor (ASM) on the {\it Ginga} satellite over the course of 4.5
years.  We describe the ASM in Section 2.  Although the ``snapshots''
of source flux taken by the ASM cannot be used to determine the
frequency histories of pulsed sources, they are a long term, uniform
set of measurements that can be correlated with other measurements.
In particular, they complement measurements of pulse frequency made
with BATSE, whose low-energy cutoff of $\sim$20\,keV misses most of
the bolometric luminosity of most pulsars, a point we return to in the
discussion.  The two instruments overlapped from 1991 April--October.
  
4U~1626--67 is a low-mass, disk-fed accreting pulsar with a 7.6\,s spin
period, a 42 minute orbital period (Middleditch et al.  1981;
Chakrabarty  et al. 1997a), and a low-mass ($M < 0.1 M_{\odot}$) helium
or carbon-oxygen dwarf companion KZ Tra (Levine et al. 1988). 
4U~1626--67 was observed to be in a state of steady spin up 
at a rate of $\dot\nu\sim 8.5\times10^{-13}$\,Hz\,s$^{-1}$ for nearly
two decades since its discovery by {\it Uhuru} in 1972.
Observations with BATSE have shown it to be in a state of spin down 
at a rate of $\dot\nu\sim -7.2\times10^{-13}$\,Hz\,s$^{-1}$ since 1991.
Quasi-Periodic oscillations in X-ray intensity with a frequency of 0.04\,Hz
were observed during spin up (Shinoda et al. 1990).
During spin down, the QPO frequency was 0.048\,Hz (Angelini et al. 1995).

\centerline{\psfig{file=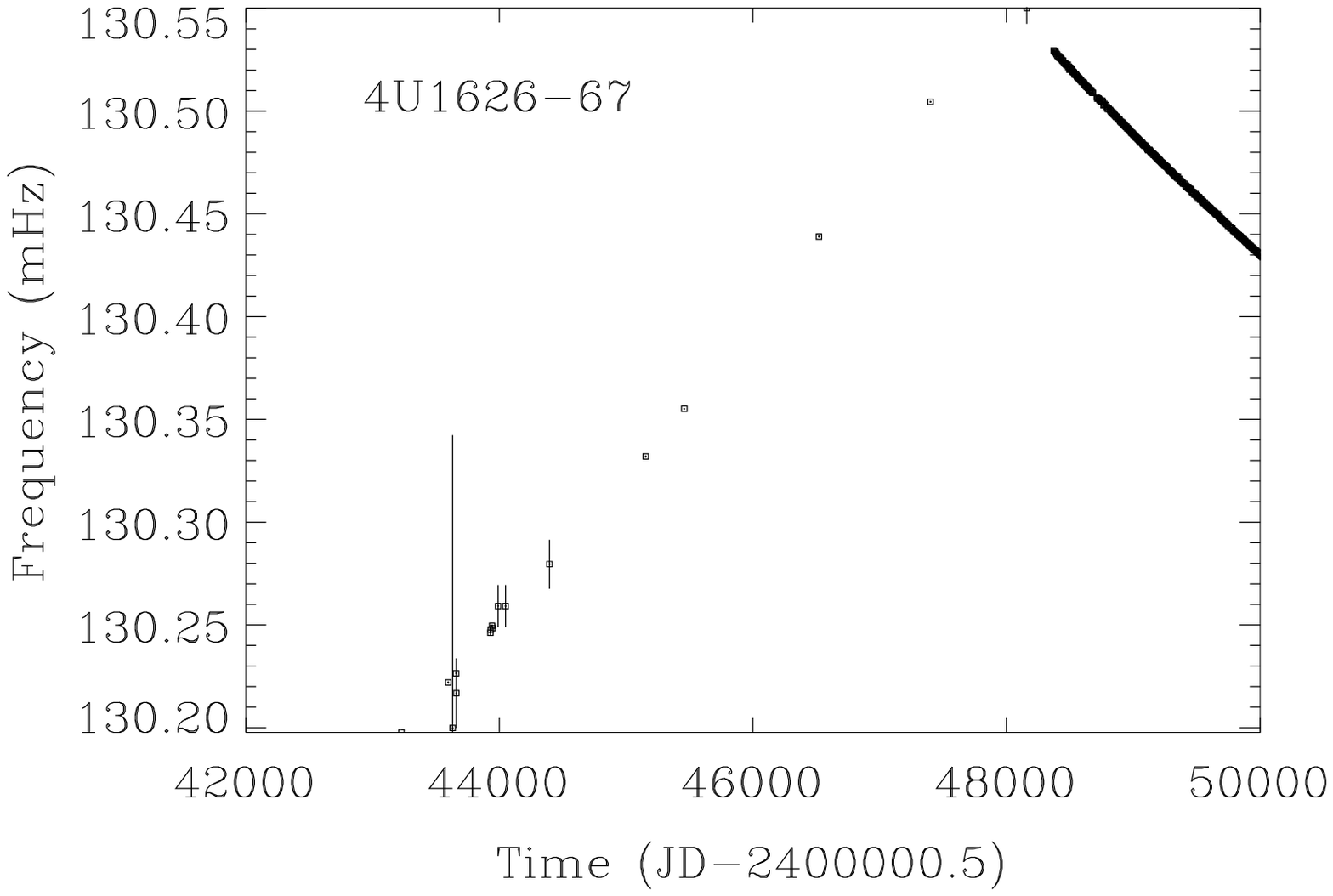,height=6cm}}
\figcaption{ The frequency history of 4U~1626--67 since 1972.}

Figure 1 shows the frequency history of 4U~1626--67.
Extrapolating the spin-up trend observed from 1975 through 1988
(Levine et al. 1988) and the spin-down trend observed by BATSE from 1991 
through the present 
(Chakrabarty et al. 1997a) yields a transition from spin-up to spin-down in
mid 1990.  We note that both the spin-up and spin-down trends have
significant higher-order components (Levine et al. 1988; Chakrabarty
1997a), hence the transition time is approximate.  The rate of
spin-down shortly after the turnaround, as measured with BATSE, 
is roughly 15\% slower than the rate of spin up prior to the turnaround
as measured by previous instruments.

GX~301-2 is a high-mass, wind fed accreting pulsar with a 680\,s spin
period, a highly-eccentric ($e=0.47$) 
42 day orbit, and a high-mass ($M\ge 40
M_{\odot}$) OB supergiant companion Wray 977 (Koh et al. 1997).  
Daily BATSE spin-frequency measurements show that most of the time,
GX~301--2 experiences a rapidly-changing accretion torque
with virtually no net change in spin frequency on long time scales.
However, BATSE observed two episodes of steady, rapid spin up from
MJD 48440--48463 (orbital phase $\phi=0.25-0.8$), and MJD 49230--49245
($\phi=0.3-0.65)$,
with an average spin-up rate of $4.5\times 10^{-12}$\,Hz\,s$^{-1}$.  
The pulsed flux in 20--55 keV during the spin-up episodes is
$1.9\times10^{-9}$\,erg\,cm$^{-2}$\,s$^{-1}$, 50\% higher
than the average pulsed flux for the same orbital phase, and almost
twice as high as the average pulsed flux over all orbital phases (Koh
et al. 1997).  The pulsed flux measured with BATSE depends strongly on
orbital phase, with a peak slightly before periastron and a secondary peak
at apastron.

\section{Observations}

The {\em Ginga\/} ASM performed well throughout the 1987 February to
1991 October period that {\em Ginga\/} was in orbit.  The effective
area of the ASM was about 420 cm$^2$, with a $45^\circ \times 1^\circ$
FWHM fan-beam collimator.  Details of the ASM appear in Tsunemi
et al. (1989).  Sky-scanning observations with the ASM were typically performed
at intervals of a few days, when the satellite was rotated
around the $z$-axis in 20 min.  During such scanning
observations, 16-channel source spectra were obtained covering the
energy range 1 to 20 keV.  An exposure time of 3--18\,s was
obtained for each scan across each observed source,
depending on the source's latitude in the spacecraft equatorial ($xy$)
plane.  For favorably located sources, the detection limit was about
50 mCrab (1--6 keV), at the 5 $\sigma$ level, worsening for sources
far from the spacecraft equatorial plane.

Data selection criteria included: (1) background low and stable, (2)
source unocculted by the Earth, and (3) acceptable spacecraft
aspect, such that the source was within $25^\circ$ of the center of
the ASM field of view.  During the 4.5 year mission, a total of 294
observations of 4U~1626--67 and 277 observations of GX~301--2
satisfied these conditions and were accepted for
follow-on analysis. 

In the discussion we make extensive use of frequencies and pulsed fluxes
measured with BATSE.  BATSE consists of 8 uncollimated
detector modules
facing outward from the corners of the CGRO spacecraft.  Each of the
8 modules contains a Large Area Detector with
a $2\pi$Sr field of view, sensitive to photons with energies of
20--1800\,keV.
Fluxes measured with
BATSE suffer from the limitation that the bulk of the bolometric flux
from most accreting pulsars is in the energy range 1--20\,keV.   Further, 
the background is large and variable.  Only
the pulsed component of the flux can be measured by epoch folding BATSE data.
BATSE observations of accreting puplsars are discussed in
detail in Bildsten et al. (1997).  

\section{Results}
\subsection{4U~1626--67}

{\it Ginga} ASM light curves of 4U~1626--67 in 1--6\,keV and 6--20 keV
are plotted in Figure 2.  Each point is the average of 30\,d
(typically $\sim$10 pointings).  The 1--6 keV count rate
shows a clear drop in mid 1990.
The average counting rate in 1--6 keV is 
0.0443(18)\,cm$^{-2}$\,s$^{-1}$ from
1987 February -- 1990 May , 
and 0.0124(37)\,cm$^{-2}$\,s$^{-1}$ from 1990 June -- 1991 November.
In 6--20 keV, the average rate is 0.0295(17)\,cm$^{-2}$\,s$^{-1}$ 
from 1987 February -- 1990 May, and 
0.0176(29)\,cm$^{-2}$\,s$^{-1}$ from 1990 June  -- 1991 November.
Count-rate variations are consistent with measurement errors within
both intervals.  Thus, the 1--6 keV count rate is smaller after 1990 June
than before 1990 June by 72\%, and the 6--20 keV count rate by 40\%.

\vskip -0.7cm
\centerline{\psfig{file=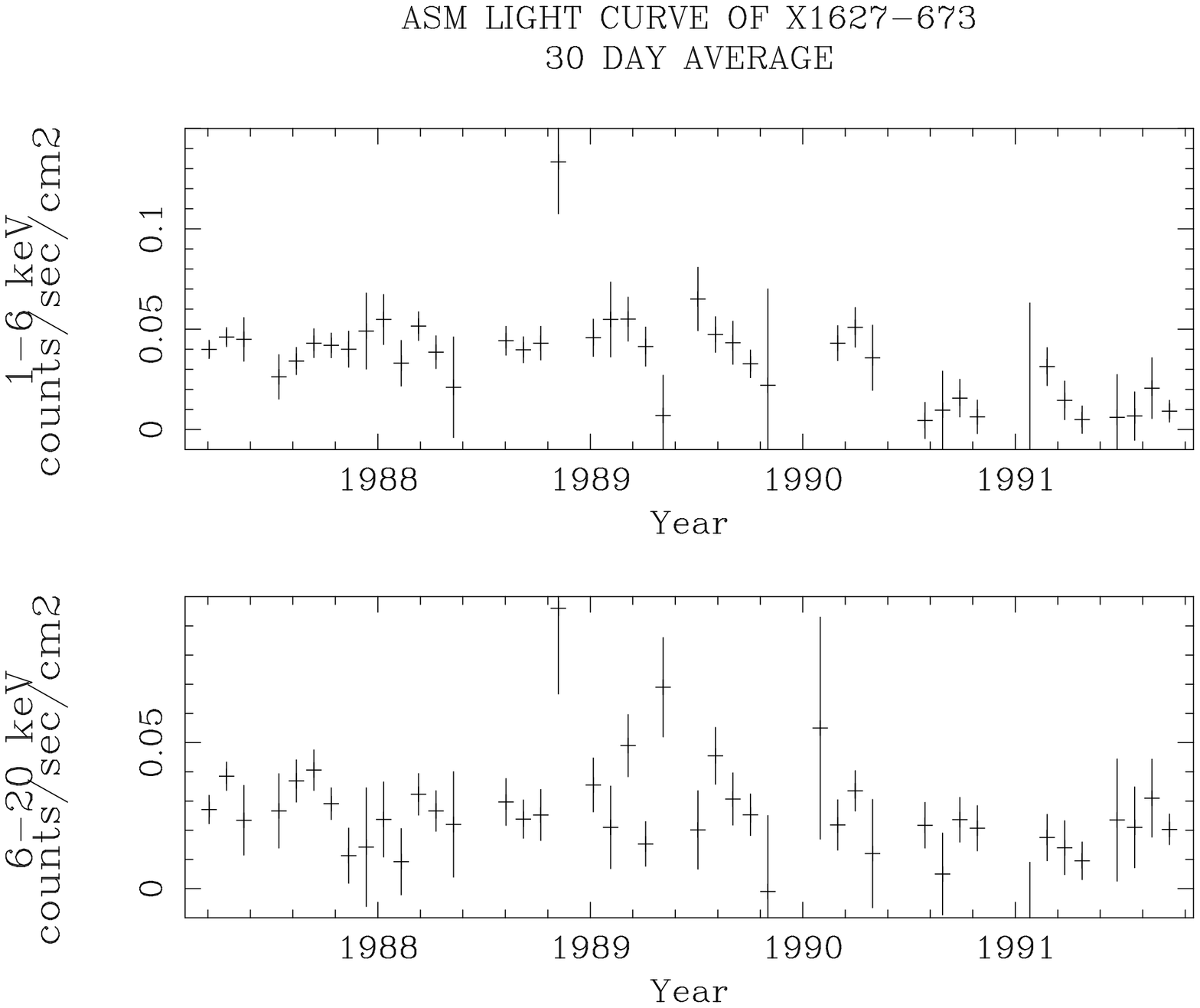,height=12cm,angle=270}}
\vskip -1cm
\figcaption{ The 4.5 year flux history of 4U~1626--67 observed
with the {\it Ginga} ASM. Each datum is an average over 30 days.}
\centerline{\psfig{file=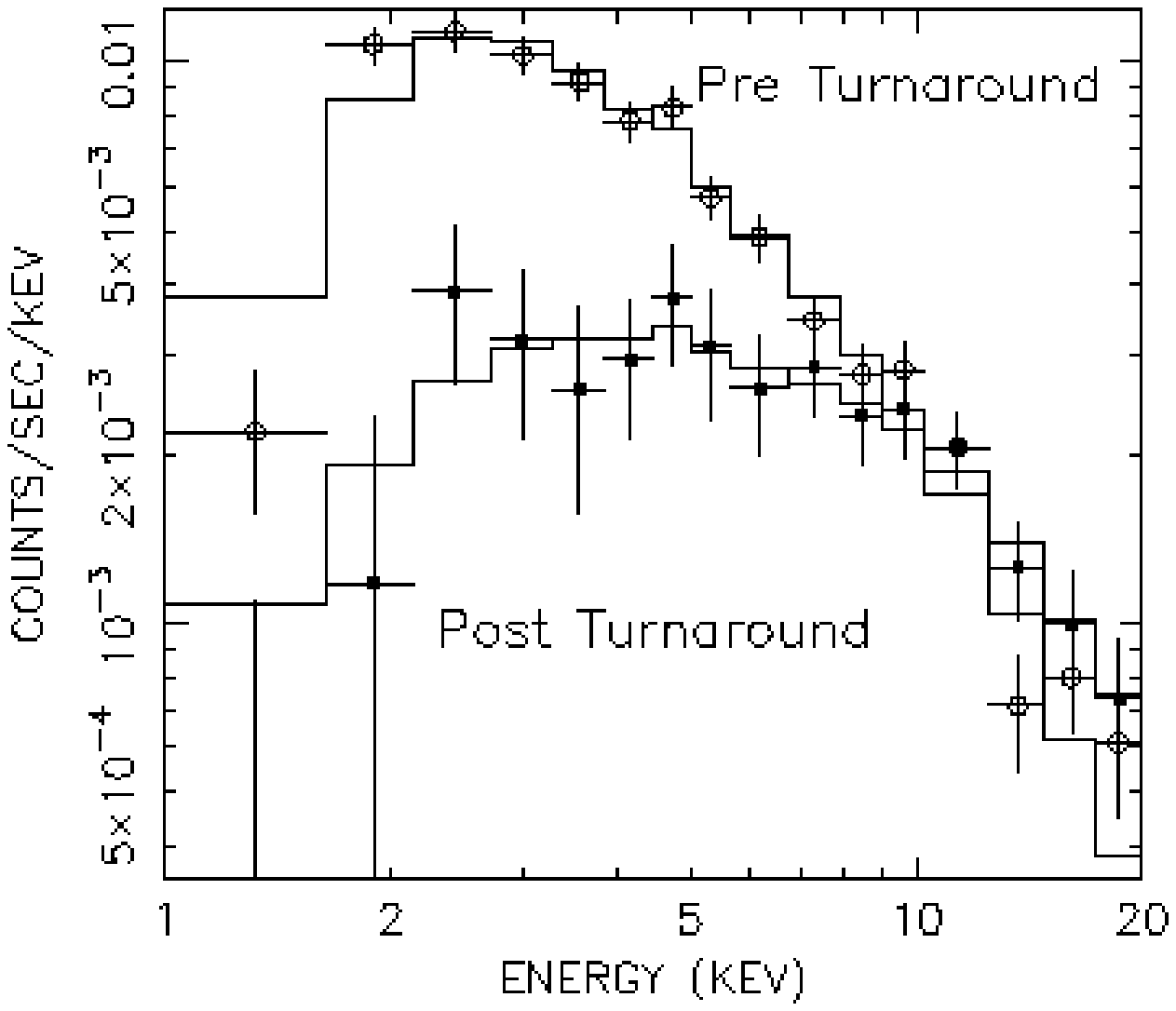,height=8cm,angle=270}}
\figcaption{
Two average energy spectra of 4U~1626--67
obtained by {\it Ginga} ASM observation.  The open circles give the 
average spectrum
before turnaround, and filled circles give the average spectrum after
turnaround.  The histograms are best-fit power law models.
}

\centerline{\psfig{file=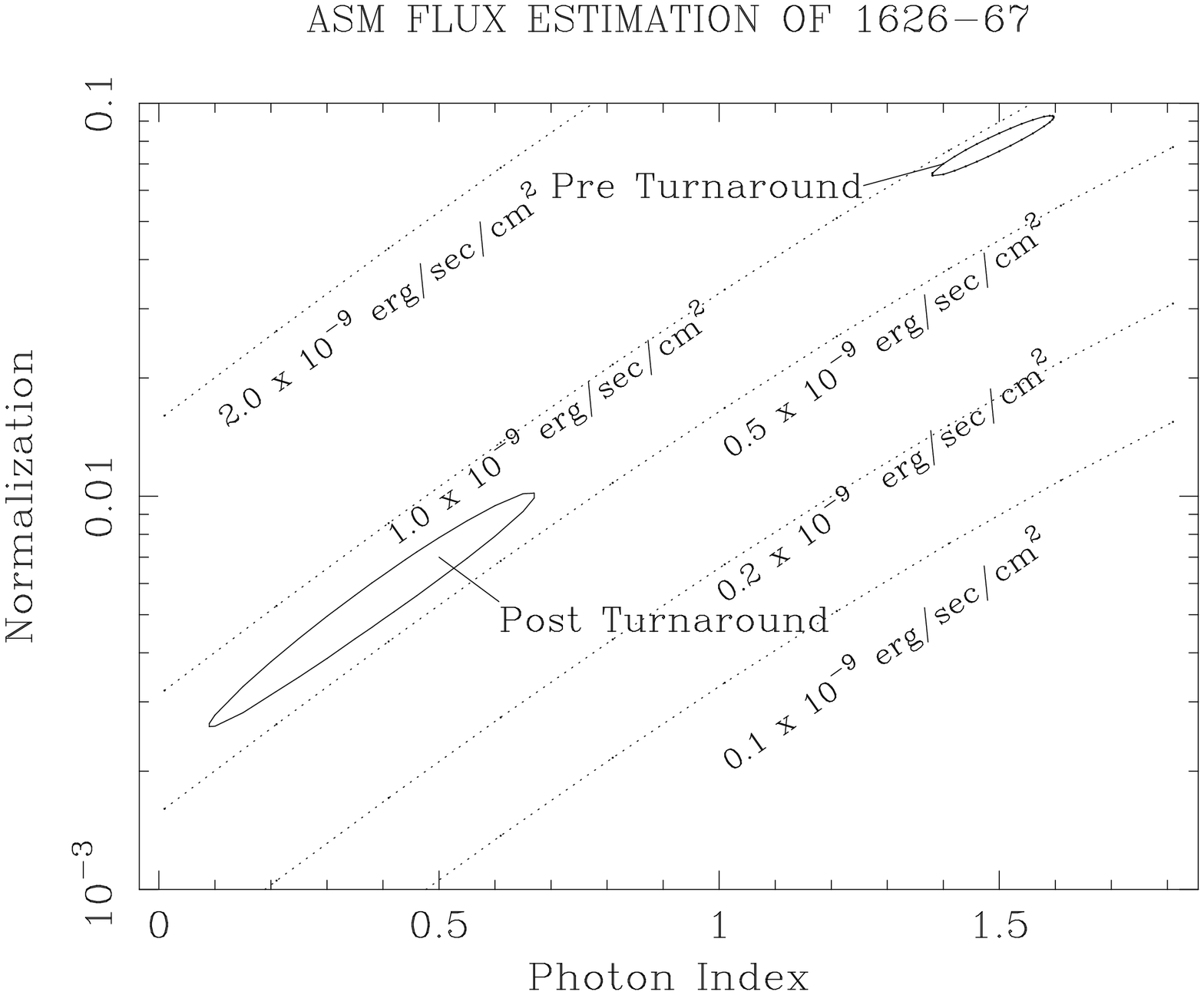,height=12cm,angle=270}}
\vskip -0.5cm
\figcaption{
Confidence contours (90
for 4U~1626--67.  Dotted lines  indicate constant flux contours.
}

Figure 3 shows the average energy spectrum of 4U~1626--67 before and
after 1990 June.  We fit both spectra by a simple power law model;
$N(E)=AE^{-\alpha}$.  Previous instruments have found the column
density to be negligible.  We verified that 
N$_{\rm H} \leq 10^{21}$\,cm$^{-2}$ 
using data obtained before and after the turnaround,
then fixed the column density at zero.  The 90\% confidence contours of
the best fit parameters are shown in Figure 4. The pre-1990 June photon
spectrum can be fit with $A=0.078(7) {\rm cm}^{-2}{\rm
s}^{-1}{\rm keV}^{-1}$ (at 1\,keV)
and $\alpha = 1.48(5)$.  The post-June
1990 spectrum can also be fit using a power law with $A=0.006(3)
{\rm cm}^{-2}{\rm s}^{-1}{\rm keV}^{-1}$ and $\alpha = 0.41(22)$.
In Figure 4, lines of constant 1-20 keV flux, $F_{\rm x}$ are ploted as
dotted lines. The flux before 1990 June is slightly higher than after.
The flux in 1--10\,keV, 10--20\,keV and 1--20\,keV are given in Table 1
for pre- and post-1990 June spectra.  The 1--10\,keV flux is observed
to drop by more than 50\%, and the 1--20\,keV flux by roughly 20\%.
The decrease in flux is less than the decrease in count rate because the
spectrum during spin down is harder.

\begin{table}[h]
\caption{4U~1626--67 X-ray Flux}
\begin{center}
\begin{tabular}{lll}\hline\hline\\[-6pt]
\multicolumn{1}{c}{Bandpass} & \multicolumn{1}{c}{Spin-up Flux $^1$} 
&
\multicolumn{1}{c}{Spin-down Flux $^1$}\\
 & \multicolumn{1}{c}{1987 Mar.--1990 May} & 
   \multicolumn{1}{c}{1990 Jun.--1991 Oct.}\\
\hline
1--20  keV & 8.89$\pm$0.56 & 6.67$\pm$0.89 \\
1--10  keV & 5.49$\pm$0.32 & 2.15$\pm$0.45 \\
10--20 keV & 3.39$\pm$0.45 & 4.48$\pm$0.80 \\
\hline
\end{tabular}
\end{center}
\noindent
$^1$ $10^{-10}$\,erg\,cm$^{-2}$\,s$^{-1}$\\
\end{table}

Chakrabarty et al. (1997a) have compared 
the spectra of 4U~1626--67 measured with a
variety of instruments including HEAO1, Einstein, {\it Ginga} LAC, and ASCA.
The photon spectral index of 0.41 measured with the {\it Ginga} ASM after
turnaround is the smallest ever measured for this source.  To evaluate
the possibility that absorption or scattering of low energy photons is
responsible for the change in spectrum, we fixed the power-law
spectral index at the pre turnaround value and attempted to fit the
post turnaround spectrum by varying the absorption.  This did not
yield an acceptable fit, and a large excess below 3 keV appears in
residuals.  A partial covering model coupled with a power law did
provide a reasonable fit, where 86(6)\% of the X-rays are obscured by
thick material with a column density of 10$^{23.8(2)}$\,cm$^{-2}$, and the
remaining X-rays are unabsorbed ($n_{\rm H}<10^{21}$\,cm$^{-2}$).  
If this partial covering model is correct, the 1--20 keV flux,
$F_{\rm x}$, is
is 1.65(47) times larger during spin down than during spin-up.

\subsection{GX~301-2}

{\it Ginga} ASM light curves of GX~301--2 in 1--6 and
6--20 keV are plotted in Figure 5. 
Each point corresponds to one scanning
observation. Since the duration of each observation (3--18\,s) is shorter
than the spin period of GX~301--2 ($\sim$680\,s), 
the count rate depends upon the pulse
phase at the time of the observation, which introduces scatter into
the measurements.

The first of the two episodes of rapid spin up observed with BATSE occurred
during {\it Ginga} operation.  Unfortunately, the {\it Ginga} ASM did not observe
GX~301--2 during the spin-up episode.
However, the periastron passage prior to spin up is well covered.
The average 6--20\,keV count rate for the four observations within 3\,d ($\pm$
0.072 in orbital phase) of the periastron passage prior to the spin-up
episode is 0.37\,cm$^{-2}$\,s$^{-1}$.
The average rate
for observations within 3\,d
of periastron, over the ASM lifetime, is
0.22\,cm$^{-2}$\,s$^{-1}$, with a standard deviation of
0.18\,cm$^{-2}$\,s$^{-1}$.
Periastron rates comparable to those seen prior to spin up
occur about every six orbits.

We constructed an average spectrum of all observations within 3\,d of 
periastron ($\phi=0.928-0.072$),
and of observations in the same range of 
phases during the periastron passage prior
to the spin-up episode, both shown in Figure 6.
We fit both with a
simple power law model with photoelectric absorption.  The 
90\% confidence contours in photon index and normalization 
are shown in Figure 7. The best fit parameters of the average periastron
passare are A = 0.204(11)\,${\rm cm}^{-2}{\rm
s}^{-1}{\rm keV}^{-1}$ (at 1\,keV),
$\alpha = 0.924(19)$, and $n_H=10^{23.61(09)}$.  
The spectral parameters of the periastron passage prior to the spin-up
episode are 
A = 1.1$^{+6}_{-0.8}$\,${\rm cm}^{-2}{\rm
s}^{-1}{\rm keV}^{-1}$ (at 1\,keV),
$\alpha = 1.2^{+0.6}_{-0.4}$, and $n_H=10^{23.83(14)}$.  

\vskip -1cm
\centerline{\psfig{file=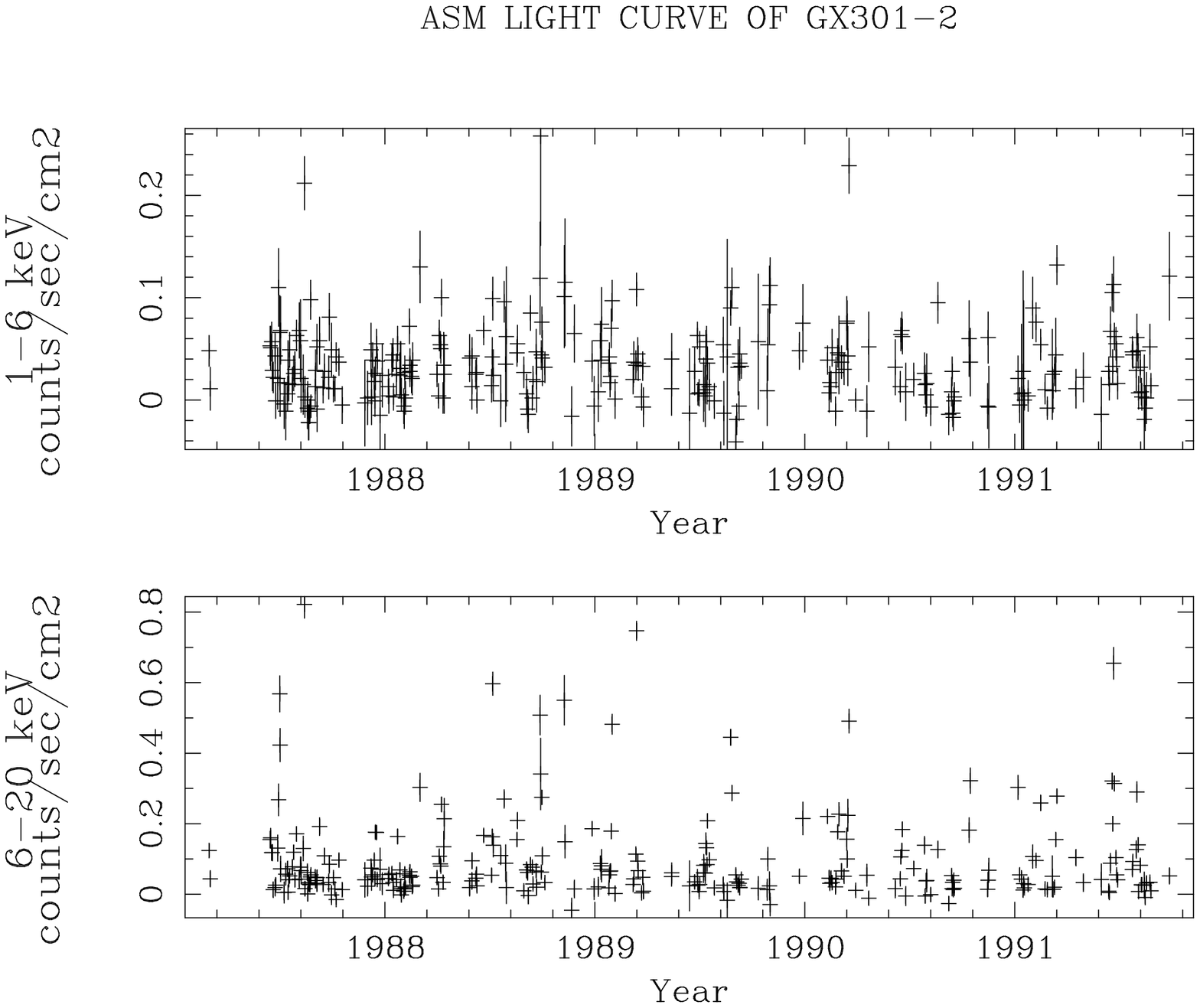,height=12cm,angle=270}}
\vskip -1cm
\figcaption{
The 4.5 year flux history of GX~301-2 observed
with {\it Ginga} ASM.  Each data point corresponds to one scaning observation.
}

\vskip -1cm
\centerline{\psfig{file=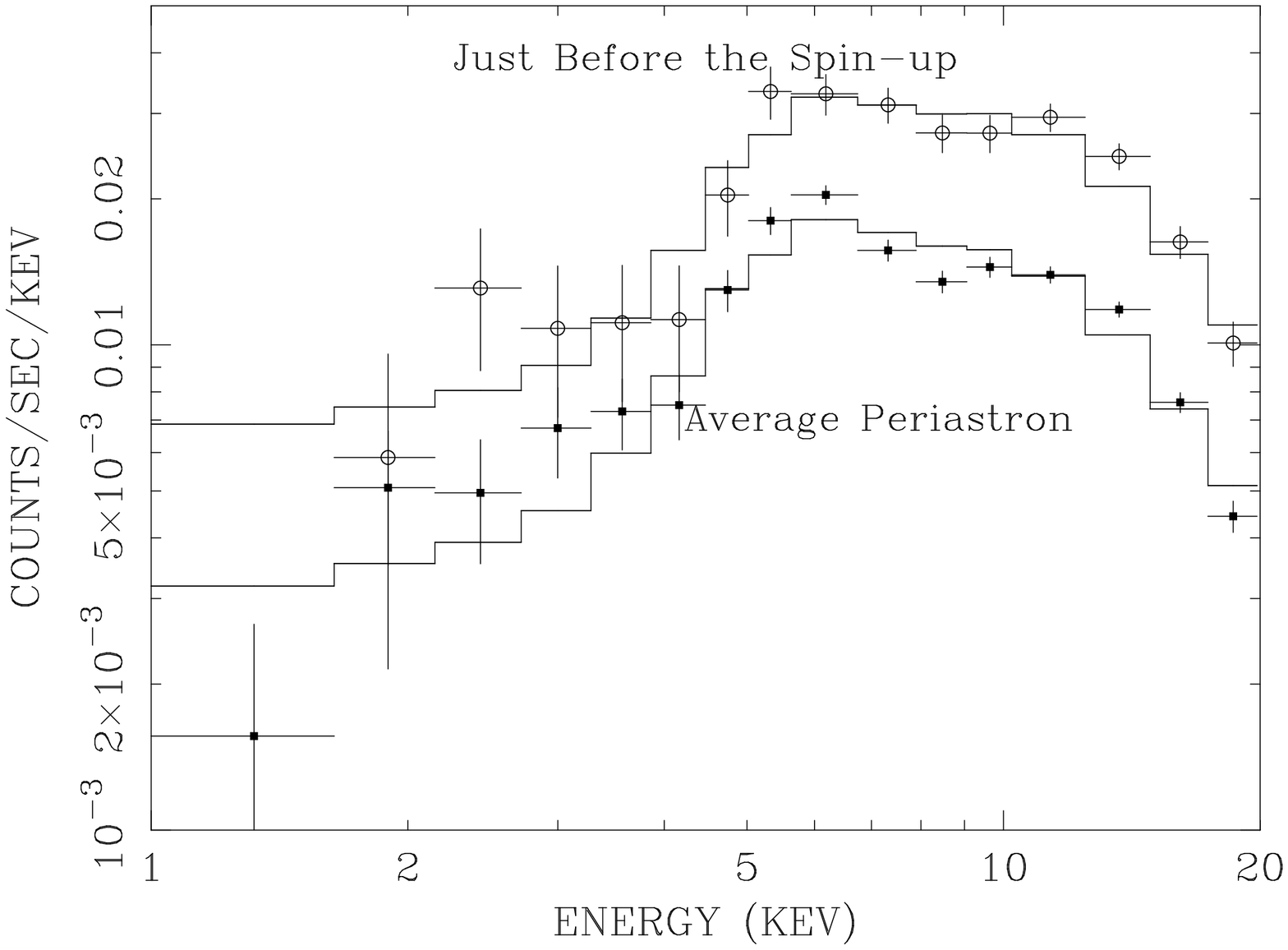,height=12cm,angle=270}}
\vskip -0.5cm
\figcaption{
Two energy spectra of GX~301--2 obtained by {\it Ginga} ASM observations.
Open circles indicate the average spectrum of four scans around the
periastron passage (orbital phase 0$\pm$0.072) just before the spin-up
episode.  The filled circles indicate the average spectrum in the
corresponding orbital phase of all {\it Ginga} ASM data.  The histograms
are best fit power law models.
}

\vskip -1cm
\centerline{\psfig{file=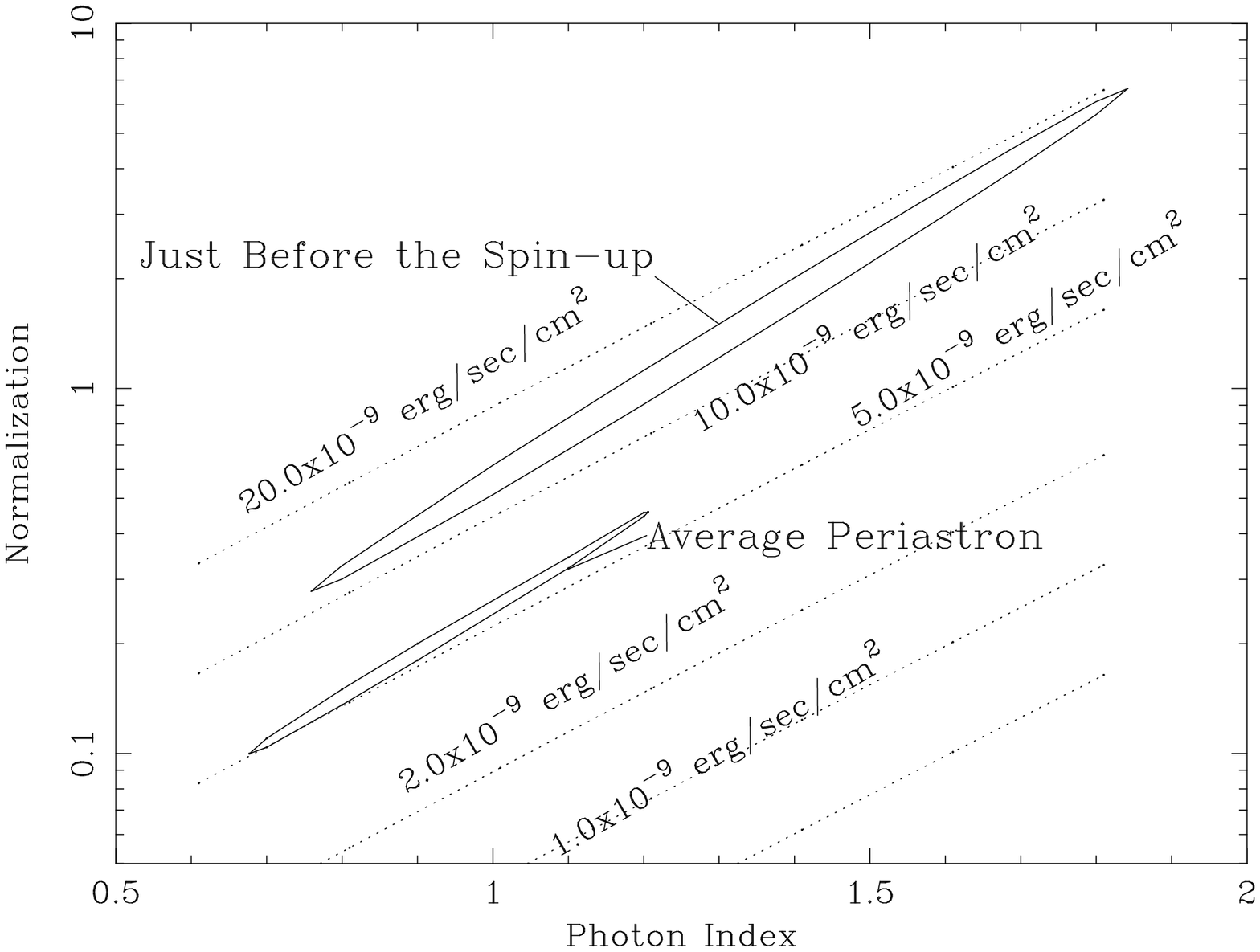,height=12cm,angle=270}}
\vskip -0.5cm
\figcaption{90\% confidence contours of the best
parameters of spectral fitting for GX301-2.  The dotted lines indicate
constant flux levels.}

\vskip 1cm
\centerline{\psfig{file=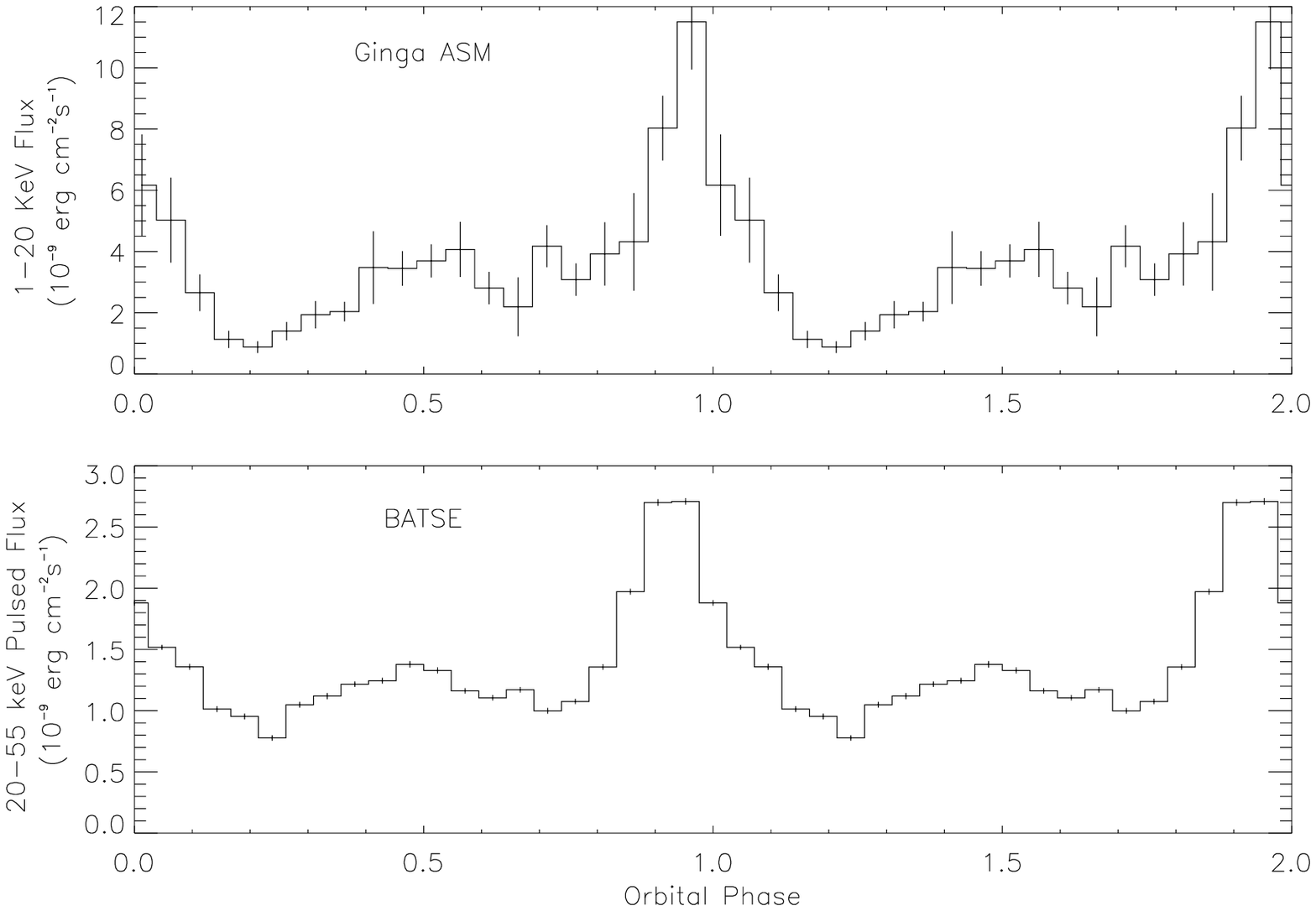,height=6cm}}
\figcaption{X-ray flux from GX 301--2 from the {\it Ginga} ASM
(top panel) and BATSE (bottom panel, from Koh et al. 1997), as a function
of orbital phase.}

The spectral shape is consistent in the two fits.
Lines of constant 6--20 keV flux
are plotted in the figure, and the 
1--20 keV and 6--20 keV fluxes are given in Table 2.
Since X-rays below 6 keV are absorbed
strongly, the uncertainty in 1--6\,keV flux is large. 

Figure 8 shows the {\it Ginga} ASM flux folded at an orbital period of
$P_{\rm orb} = 41.488$\,d and referenced to an epoch of periastron 
of MJD 48,802.85 (Koh et al. 1997), along with the BATSE pulsed flux
folded with the same orbit (Koh et al. 1997).  
The {\it Ginga} ASM fluxes were determined by folding 
ASM count rates corrected for effective area, then converting to a 
flux by assuming a constant spectral shape.  The {\it Ginga} ASM and BATSE
fluxes both show a peak shortly before periastron, a broad secondary
maximum at apastron, and a minimum at orbital phase $\sim$0.2.  The
1--20\,keV flux is a factor of $\sim$3 larger than the 20--55\,keV pulsed
flux, and shows amplitude modulations a factor of $\sim$2 larger.

\begin{table*}[b]
\caption{GX 301-2 X-ray Flux}
\begin{center}
\begin{tabular}{lll}\hline\hline\\[-6pt]
\multicolumn{1}{c}{Bandpass} & \multicolumn{1}{c}{Spin-up Episode$^{*}$} 
&
\multicolumn{1}{c}{Average Periastron$^{*}$}\\
 & \multicolumn{1}{c}{4 data in $\pm$ 3 days} & 
   \multicolumn{1}{c}{$\pm$ 3 days}\\
\hline
1--20 keV & 27.46$\pm$13.24 & 7.64$\pm$1.49 \\
6--20 keV & 15.11$\pm$3.94 & 5.60$\pm$0.52 \\
\hline
\noindent
$^{*}$ $10^{-9}$\,erg\,cm$^{-2}$\,s$^{-1}$
\end{tabular}
\end{center}
\end{table*}

\section{Discussion}

Dynamical tests of accretion theory require simultaneous 
measurements of torque and bolometric luminosity.  The observations
presented here provide indirect evidence for a connection between
accretion torque and luminosity in the persistent accreting pulsars
4U~1626--67 and GX~301--2.

The 1--20\,keV flux is 
20\% lower in 4U~1626--67 during
spin down than during spin up.  
No other single instrument observed the
source in both states.  
It is notable that the 1--10\,keV flux
decreased by 55\%, compared with a 20\% change in the 1--20\,keV flux.
Measurements with the soft X-ray instruments
ROSAT and ASCA during spin-down both yielded
decreases of 60\% or larger relative to the flux in the same band
measured in 1979 with
HEAO.  The change in luminosity measured with the ASM is close
to the 15\% change in the magnitude of the accretion torque measured
with BATSE.
We now consider implications of our measurement to models of accretion torque.
In the following discussion we will sometimes denote the spin frequency of the
neutron star $\nu_*$, rather than $\nu$, to avoid confusion with other frequencies.

Pulsars cannot spin down while accreting in 
the simple picture of disk accretion outlined in \S 1. 
Near equilibrium, however, the accretion torque may depend in detail upon the
disk-magnetosphere interaction.  Spin-down can occur while material
continues to accrete if negative, ``non-material'' torques are present.
These may result from dragging of the magnetosphere by the disk outside 
the corotation radius, where the disk rotates more slowly than
the neutron star (Ghosh \& Lamb 1979).  Alternatively,
mass ejection in a magnetohydrodynamic wind may slow
the neutron star (Arons \& Lea 1980).  Both models predict that a drop in
$\dot{M}$ causes a decrease in $N$, and potentially a reversal from
spin up to spin down.  

  The model of Ghosh \& Lamb is the most detailed.  They derived a modified 
torque equation given by $N = n(\omega)\dot M (GM_{\rm x}r_{\rm m})^{1/2}$.
The dimensionless function $n(\omega)$ is the same for all accreting pulsars.
The ``fastness'' is defined as $\omega \equiv \nu_*/\nu_K(r_{\rm m})$, and is 
proportional to $\dot M^{-3/7}$.  The function $n(\omega)$ increases smoothly
with $\dot M$.  It passes through zero at a critical value 
$\omega=\omega_c$, and approaches $n(\omega)=1$ when $r_{\rm m}\ll r_{\rm co}$.
Its exact form is somewhat controversial, but it is sufficient
for our purposes that it can be written approximately as

\begin{equation}
n(\omega)\approx N_0\left({{1-\omega/\omega_c} \over {1-\omega}}\right)
\label{GL}
\end{equation}

If the model of Ghosh and Lamb is correct,
we can determine how close 4U~1626--67 is to equilibrium.
In going from spin up
to spin down, the change in accretion torque was $\delta N / N \sim -2$.
Assuming $\dot M \propto L_{\rm x}$, the change in $\dot M$ was
$\delta \dot M / \dot M = -0.2$.  In general,
$\delta \omega / \omega = -3/7 (\delta \dot M / \dot M)$ and
$\delta N / N = (-\delta\omega / \omega_c) / (1 - \omega / \omega_c) +
 \delta \omega / (1 - \omega) + 6/7 ~\delta \dot M/\dot M$.
Near equilibrium,
where $\omega\approx\omega_c$ and $1-\omega/\omega_c\ll 1-\omega$, we have
$\delta N / N \approx (\delta \omega / \omega) / (1 - \omega / \omega_c) 
+ 6/7~  \delta \dot M/\dot M =
3/7~ (\delta \dot M / \dot M) 
[1 / (1 - \omega / \omega_c)+2]$.  From the measured
$\delta \dot M / \dot M$ and $\delta N / N$ we find 
that during spin up, $w \approx 0.95~\omega_c$.

We can constrain the distance, $D$, to 4U~1626--67, from the 
luminosity and flux during spin up and from 
$\omega \approx 0.95~\omega_c$.
Combining equations \ref{N} and \ref{Lx} and using $N=2\pi I_{\rm x}\dot\nu$
yields 
$2\pi I_{\rm x}\dot\nu = r_{\rm x}L_{\rm x}(GM_{\rm x})^{-1/2}r_{\rm m}^{1/2}$.
Solving for $r_{\rm m}$ in terms of $r_{\rm co}$ and $\omega_c$ and using
$r_{\rm co}^3=GM_{\rm x}P_{\rm spin}^2/(4\pi^2)$ then gives us
$L_{\rm x}=(2\pi)^{4/3}(0.95\omega_c)^{-3/4}I_{\rm x}\dot\nu r_{\rm x}^{-1}
           (GM_{\rm x})^{1/3}P_{\rm spin}^{-1/3}$.
Finally, we assume that the 1--20 keV flux, 
$F_{\rm x}$, is a fraction, $1/\eta$, of the bolometric
flux.  Using $\eta F_{\rm x}=L_{\rm x} / (4\pi D^2)$, and putting in the
measured values of $F_{\rm x}$, $\dot\nu$ and $P_{\rm spin}$ during
spin up yields

\begin{equation}
\left({D \over {\rm 5\,kpc}}\right)^2
\left({{r_{\rm x}} \over {\rm 10\,km}}\right)
\left({{M_{\rm x}} \over {1.4 M_\odot}}\right)^{-1/3}
\left({I_{\rm x} \over {10^{45}{\rm g\,cm}^2}}\right) = 
{1\over{\eta\sqrt{\omega_c}}}
\label{D}
\end{equation}

$D$ must be at least $\sim$5\,kpc for a neutron-star with a mass of 
$1.4 M_\odot$, a radius of 10\,km and a moment of intertia of
$10^{45}$\,g\,cm$^2$.  Chakrabarty et al. 1997a obtained a lower limit of
3\,kpc using the higher flux value of 
$2.4\times10^9$\,ergs\,cm$^{-2}$\,s$^{-1}$ measured with HEAO1 
(Pravdo et al. 1979).

  We can go on to determine $\omega_c$ if we further assume that the
QPO observed in 4U~1626--67 during spin up is a magnetospheric beat-frequency
oscillation satisfying $\nu_{\rm QPO}=\nu_K(r_{\rm m})-\nu_*$ (Alpar \& Shaham
1985; Lamb et al. 1985).  By definition, $\omega = \nu_*/\nu_K(r_{\rm m})$.
Substituting $\nu_K(r_{\rm m})=\nu_{\rm QPO}-\nu_*$, with 
$\nu_*=0.13$\,Hz and $\nu_{\rm QPO}=0.04$\,Hz, and using 
$\omega=0.95\omega_c$, yields $\omega_c=0.8$.  Equation~(\ref{D}) then
yields a distance of $\sim$5\,kpc for $\eta$ of order unity.

If $\nu_{\rm QPO}=\nu_K(r_{\rm m})-\nu_*$, 
we would have expected $\nu_{\rm QPO}$ to 
decrease in going from spin up to spin down.  
In fact the QPO frequency increased from 0.04\,Hz
to 0.048\,Hz.
One possible explanation is that
$\nu_K(r_{\rm m})$ decreased from $\sim$0.17\,Hz to $\sim$0.08\,Hz and
that $\nu_{\rm QPO}$ changed from 
$\nu_K(r_{\rm m})-\nu_*$ to $\nu_*-\nu_K(r_{\rm m})$.
This interpretation has two problems.  First, it requires
the magnetospheric radius to move outside the corotation
radius.  It is not known how accretion can occur when
$r_{\rm m}>r_{\rm co}$.  Second, the change in $\nu_K(r_{\rm m})$ then
requires an approximately
80\% decrease in $\dot M$.  In contrast, 
we observed a 20\% decrease in $L_{\rm x}$.  
It is unlikely that the
dependence of $r_{\rm m}$ on $\dot M$ changes significantly
near equilibrium since the magnetic energy density is a such strong function
of radius ($B^2\propto r^{-6}$).  We think it more likely that the QPO in 
4U~1626--67 is not a magnetospheric beat-frequency oscillation.  If not,
then we cannot estimate $\omega_c$ from these observations, and 
equation~(\ref{D}) provides only a lower limit to the distance.

A partial covering model provides a reasonable fit to 
the energy spectrum of 4U~1626--67 for
1990 June -- 1991 October, with a 60\% higher X-ray flux
than 1987 April -- 1990 June.
Observationally, partial covering cannot be ruled out.  
However, the flux from  4U~1626--67 is low enough that even for a distance
of 10\,kpc the luminosity is substantially sub Eddington.  We 
do not expect the physical conditions
associated with partial absorption in wind-fed systems like as Vela X-1, 
such as an accretion wake, to be present.  

In wind-fed systems small, transient accretion disks with frequent
reversals in rotational sense
are thought to form (e.g. Wang et al. 1981; Fryxell \& Taam
1988).  Wind-fed pulsars, such as Vela~X-1 and GX~301--2, display erratic
spin-frequency behavior that can be described as a random walk in spin
frequency (Deeter \& Boynton 1982; Bildsten et al. 1997).  
Because neither the rate nor the
sense of accretion is constant, equilibrium is not a meaningful
concept to apply to wind accretors.  Correlations between torque and
luminosity would have to be measured on the time scale of disk formation 
and reversal, which is thought to be of order hours to days.

The two episodes of steady spin up observed with BATSE in GX~301--2 were
probably associated with the formation and accretion of a disk.  During
both episodes, the 20--55\,keV pulsed flux was $\sim$50\% higher than average.
However, since the average 1--20\,keV flux
is 3 times larger than the average 20--55\,keV pulsed flux 
(see Figure 8),
a change in spectral shape or pulsed fraction could have caused the
enhancement seen by BATSE.
The {\it Ginga} ASM did not observe GX~301--2 during the spin-up episode.
However, the {\it Ginga} ASM flux, folded at the orbital period, is in good
agreement with the BATSE pulsed flux folded with the same orbit.
This agreement makes it unlikely that the increase of pulsed
flux during the spin-up episode 
resulted just from a change in spectral shape or pulsed fraction, and
strongly supports the correlation between luminosity and accretion
torque inferred from the BATSE measurements.  The 
flux enhancement seen with the {\it Ginga} ASM during the periastron 
passage prior to spin-up
is further evidence for an increase in total flux during spinup.
On the other hand, 
such an enhancement is seen in other periastron passages, and
no such enhancement
in the 20--55\,keV pulsed flux is seen with BATSE during the periastron 
passage prior to spin up relative to other periastron passages.

Chichkov et al. see a periastron flare in the 1--20\,keV flux with 
WATCH/GRANAT, along with evidence of flux enhancement at apastron and
other phases.  They measure an average peak periastron luminosity
of $1.4\times10^{37}$ergs\,s$^{-1}$ for a distance of $D=1.8$\,kpc,
a factor of 2 larger than the flux from the {\it Ginga} ASM.
Their larger flux
may be due to the harder spectral shape they assumed
in converting from count rate to flux (White, Swank, \& Holt 1983).

Our measurements provide indirect evidence for a connection between
torque and luminosity in the persistent accreting pulsars
4U~1626--67 and GX~301--2.
They reveal a $\sim$20\% change in the 1--20\,keV X-ray
flux of 4U~1626--67 between spin up and spin down, comparable to
the change in $|\dot\nu|$.  The ASM monitored 4U~1626--67 during the
last 3 years of its extended spin up episode and during the
first 16\,months of spin down.
The energy spectrum is significantly harder during spin down
than during spin up -- the change in the 1--10\,keV flux is $\sim$50\%.
These observations highlight the importance of continuous, broad-band
monitoring.

We confirm that the X-ray flux of GX~301--2
varies with orbital phase, as observed in 20--55\,keV pulsed flux
with BATSE.  The shape of the 1--20\,keV X-ray light curve shows all 
the features of the BATSE light curve.  Although we do not observe
the 1991 episode where BATSE observes an enhancement in the 
20--55\,keV pulsed flux accompanied by smooth, rapid spin up, the
close similarity between the ASM and BATSE light curves indicate that
the BATSE pulsed flux traces the bolometric flux, supporting the
correlation between torque and luminosity suggested by the BATSE
observations.
The details of the connection between torque and luminosity in 
persistent pulsars remain
unknown, and will require continued, simultaneous monitoring of the
X-ray flux and spin frequency of these objects.

\acknowledgements{We wish to thank Robert Nelson, Danny Koh, and Deepto
Chakrabarty for helpful comments and discussion.  B.A.V. acklowledges
support from the United States National Aeronautics and Space Administration
under grants NAG5--3293 and NAG5--1458, and from the United States National
Science Foundation under grant INT950--3950.  
This work was partially supported by Grant-in Aids from the Japanese Ministry
of Education, Science, Sports and Culture (No. 09640320).}

\clearpage

\end{document}